\documentclass[12pt]{iopart}

\usepackage{graphicx}
\usepackage{caption}
\usepackage{subcaption}
\usepackage{hyperref}
\usepackage{colortbl}
\usepackage{xcolor}
\pagecolor{white}

\begin{document}

\title[Fast Neural Network Inference on FPGAs for Triggering on LLPs at Colliders]{Fast Neural Network Inference on FPGAs for Triggering on Long-Lived Particles at Colliders}

\author{Andrea Coccaro$^{1}$, Francesco Armando Di Bello$^{1,2}$, Stefano Giagu$^{3}$, Lucrezia Rambelli$^{1,2}$ and Nicola Stocchetti$^{3}$}
\address{$^{1}$INFN, Sezione di Genova, 16146 Genova, Italy}
\address{$^{2}$Department of Physics, Università di Genova, 16146 Genova, Italy}
\address{$^{3}$Department of Physics, Università La Sapienza and INFN Sezione di Roma, 00185 Rome, Italy}

\ead{andrea.coccaro@ge.infn.it}

\vspace{10pt}

\begin{abstract}
Experimental particle physics demands a sophisticated trigger and acquisition system capable to efficiently retain the collisions of interest for further investigation. Heterogeneous computing with the employment of FPGA cards may emerge as a trending technology for the triggering strategy of the upcoming high-luminosity program of the Large Hadron Collider at CERN. In this context, we present two machine-learning algorithms for selecting events where neutral long-lived particles decay within the detector volume studying their accuracy and inference time when accelerated on commercially available Xilinx FPGA accelerator cards. The inference time is also confronted with a CPU- and GPU-based hardware setup. 
The proposed new algorithms are proven efficient for the considered benchmark physics scenario and their accuracy is found to not degrade when accelerated on the FPGA cards. The results indicate that all tested architectures fit within the latency requirements of a second-level trigger farm and that exploiting accelerator technologies for real-time processing of particle-physics collisions is a promising research field that deserves additional investigations, in particular with machine-learning models with a large number of trainable parameters.
\end{abstract}

\vspace{2pc}

\section{Introduction}
A crucial aspect of particle physics experiments at colliders is the trigger and data acquisition system.
In fact, efficiently collecting the products of the collisions resulting in interesting physics processes is a challenging task, for both the complexity and sparsity of the detector data to be analysed and the stringent latency requirements imposed by the high frequency of the occurring collisions.

Both the ATLAS and CMS experiments~\cite{ATLAS-Paper,CMS-Paper}, being the two multi-purpose particle-physics detectors with cylindrical geometry currently running at the Large Hadron Collider (LHC) at CERN~\cite{LHC}, employ a two-tier trigger system for selecting the products of the proton-proton collisions, so-called events, for storage and analyses~\cite{ATLAS-Trigger,CMS-Trigger}.  

The initial 40 MHz rate of proton-proton collisions produced by the LHC is first reduced to~$\mathcal{O}(100 \mbox{ kHz})$ by a hardware-based Level-1 (L1) trigger system, and then further reduced down to~$\mathcal{O}(1 \mbox{ kHz})$ by a software High Level Trigger (HLT). Triggering events is therefore an optimisation problem: how to maximise the variety and richness of the physics program with the limitations in terms of latency, throughput, data transfer, and storage capabilities. The selection at L1 must occur with a latency of~$\mathcal{O}(10^{-1}\div 10^0\,\mu \mbox{s})$ and is obtained by using low-resolution detector information. The selection at the HLT, instead, is based on software running on a commercial CPU-based farm, and, with access to more granular detector information, needs to occur with typical latency times between~$\mathcal{O}(10^{-1} \div 10^{0}\,\mbox{s})$.

With the upcoming high-luminosity phase of the LHC (HL-LHC)~\cite{HL-LHC}, the design of the trigger and data acquisition system needs to cope with the higher occupancy and the higher number of readout channels of the upgraded detectors. The advancement in single-processor computing performance is not adequate, and more modern solutions of heterogeneous computing may offer an interesting avenue of exploration~\cite{ATLAS-HL-LHC-Computing,CMS-HL-LHC-Computing}. In particular the works presented in~\cite{Duarte:2019fta,Rankin:2020usv} suggest that FPGA-accelerated inference of machine-learning algorithms is a promising option for particle physice experiments, requiring minimal modifications to the current computing models.

In this context, we study the possibility to implement algorithms based on deep neural networks for the event selection at the HLT, and to use commercial accelerator boards based on FPGA processors to improve the performance in terms of processing time and throughput.
FPGAs are reconfigurable hardware architectures which can be adapted for specific tasks and are traditionally programmed using hardware description languages like VHDL or Verilog. In recent years several tools and libraries were developed to facilitate the implementation and deployment of both traditional and machine learning algorithms on FPGAs. The Xilinx~\cite{Xilinx} company for example has released Vitis-AI~x\cite{Vitis-AI}, being an AI-inference development platform for AMD devices, boards, and Alveo data center acceleration cards. Similarly, Intel has developed the FPGA AI Suite based on OpenVINO~\cite{OpenVINO}.

In this work we construct and characterize deep neural networks targeting the selection of events where neutral long-lived particles decay within the detector volume. We present the design and the results of the implementation in a working engineering pipeline that starts from the pre-processing of the input data, to the training of the deep neural network-based model, to the optimization and deployment on two Xilinx FPGA accelerators, the Alveo U50 and the Alveo U250, all based on the use of publicly available libraries. Two approaches based on a deep convolutional neural network and on an autoencoder are developed and presented. A comparison of the performances of the deployed algorithms in CPU, GPU and FPGA accelerators is also shown. We stress the complementarity of this approach, also in terms of development and maintenance of the needed libraries, with respect to the ongoing work of deploying neural networks on FPGA boards with a latency compatible with the selection occurring at L1, where a dedicated software library, \texttt{hls4ml}, is being developed~\cite{Loncar:2020hqp,Aarrestad:2021zos}, and dedicated implementations have been recently proposed~\cite{Francescato:2021mca}.
The paper is organized as follows. In Section~\ref{Sec:Physics} we describe the physics benchmark and the dataset. In Section~\ref{Sec:Models} we introduce the trigger strategies we have tested, and the associated algorithms: a convolutional neural network (CNN) and an autoencoder (AE) architecture. In Section~\ref{Sec:Results} we present and discuss the results. Finally, we provide our concluding remarks in Section~\ref{Sec:Conclusions}.

The dataset used for the presented results is made available in Zenodo at the link in Ref.~\cite{Zenodo}. The codes for constructing the algorithms, converting the models and evaluating their performances are available on request by contacting the authors.

\section{Physics benchmark and datasets}\label{Sec:Physics}
The Standard Model (SM) of particle physics provides an excellent description of all observed phenomena up to the energies presently explored. A variety of beyond the SM scenarios has been proposed in literature to address open questions such as naturalness, baryogenesis, dark matter and the origin of neutrino masses, and new long-lived particles (LLPs) are often present in these scenarios~\cite{Alimena:2019zri}.

Among the models with neutral LLPs, the ones in Ref.~\cite{Strassler:2006im,Strassler:2006ri,Falkowski:2010cm,Falkowski:2010gv} are of particular interest because of the predicted unconventional phenomenology in the collisions at the LHC. The peculiarity of the reconstructed final states, containing collimated signatures of pairs of leptons and/or light hadrons, with no detector activity connecting such signatures with the interaction point of the colliding protons, resulted in the development of dedicated signature-driven triggers to enable the searching of these models at the LHC and HL-LHC~\cite{ATLAS:2013bsk,Coccaro:2016lnz,Bhattacherjee:2019fpt,Bhattacherjee:2020nno}.

This paper focuses on the identification of a neutral LLP decay with the data collected by the muon spectrometer (MS) of a typical experiment at the LHC. In this way, the the search sensitivity, both in terms of geometrical acceptance and of reduced backgrounds, is typically maximised. A toy simulation of the monitored drift tube (MDT) detector together with the superconducting toroidal magnetic field of the ATLAS experiment is developed, together with the physics benchmark of a neutral LLP decaying to charged particles. In particular the generation, simulation and reconstruction chain can be summarised as follows:
\begin{enumerate}
\item generation of a neutral LLP and of its charged decay products in the MDT detector volume and within the magnetic field;
\item simulation of the detection of the decay products through the formation of hits in the MDT chambers;
\item estimation of the experimental effects on the hit positions using the resolutions of the ATLAS MDT detector as in~\cite{Diehl:2009bw};
\item addition of detector noise and background accounting for the measured average rate during the data-taking of the LHC~\cite{ATLAS:2022jjr}. 
\end{enumerate}
The simulated experimental conditions are considered to be of enough detail for the scope of this article, which is to demonstrate, as a proof of principle, the benefits of inference acceleration in the context of triggering applications for particle-physics experiments.

\begin{figure}
\centering
\begin{subfigure}{0.49\textwidth}
\centering
\includegraphics[width=7.5cm]{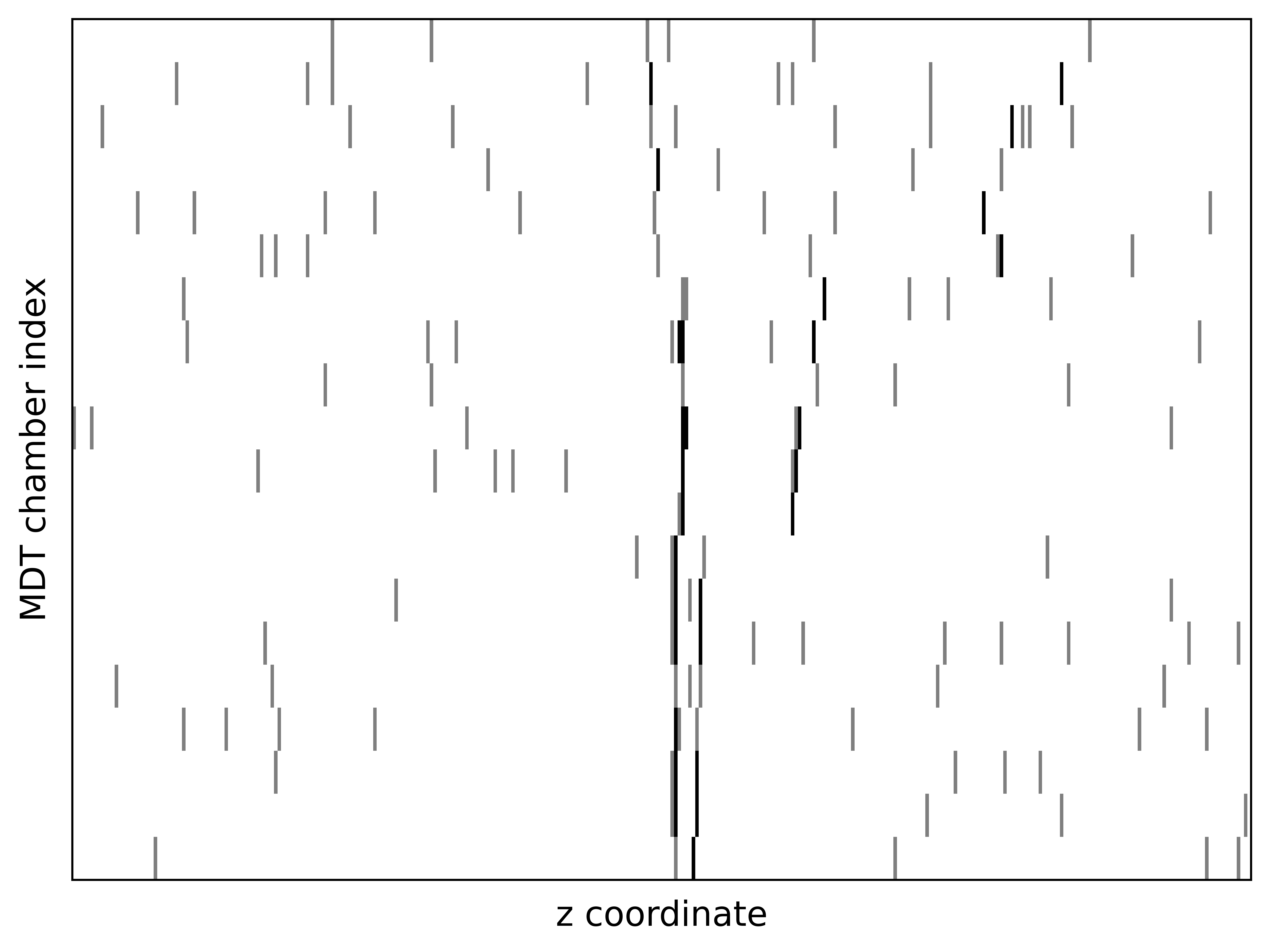}
\caption{Two-track LLP decay}
\end{subfigure}
\hfill
\begin{subfigure}{0.49\textwidth}
\centering
\includegraphics[width=7.5cm]{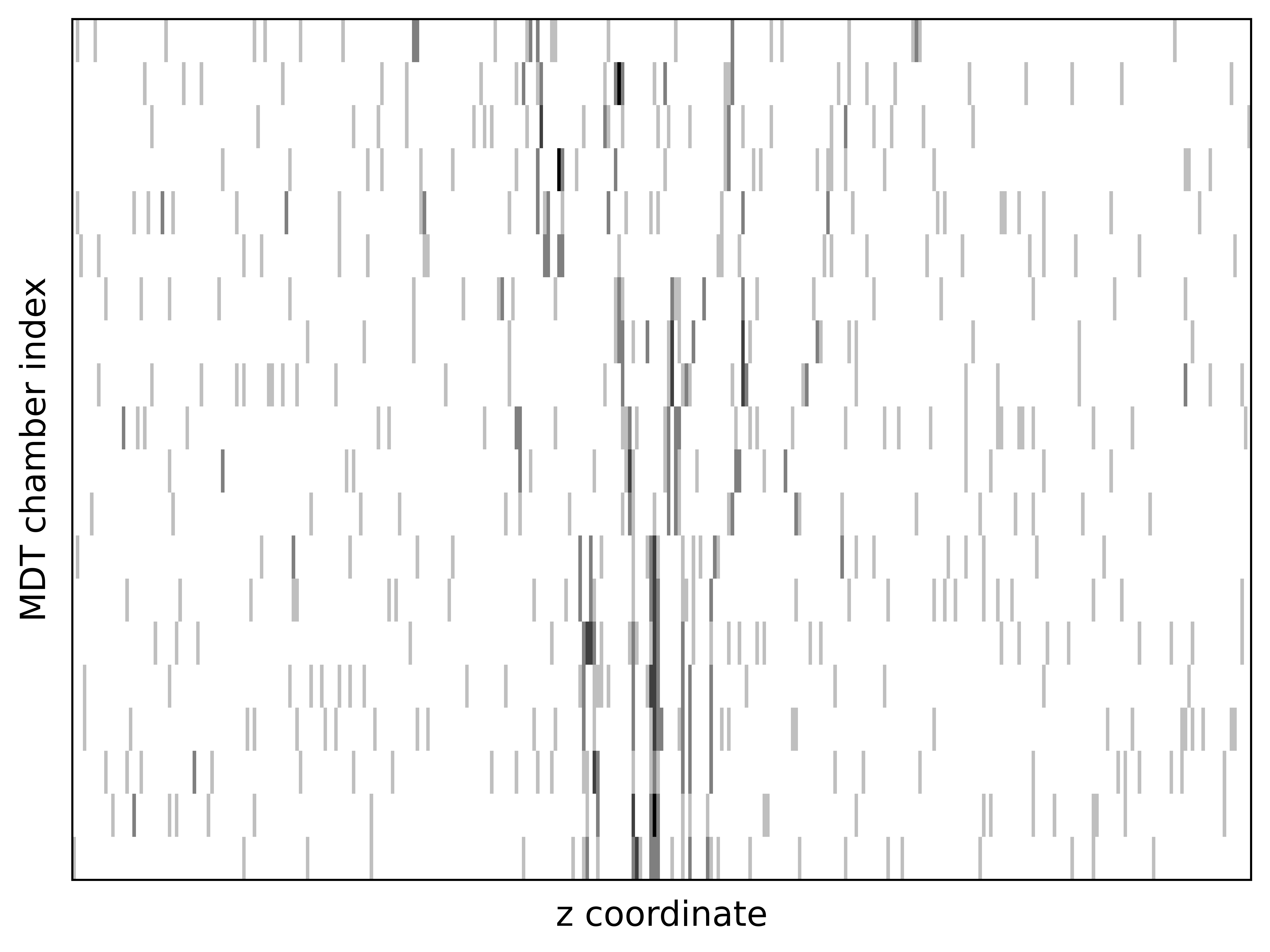}
\caption{Ten-track LLP decay}
\end{subfigure}
\caption{Image representation of a neutral LLP decaying to two (a) and ten (b) charged particles with the signal pattern and detector noise released into the MDT chambers.}
\label{fig:events}
\end{figure}

Physics processes are simulated with a number of charged particles as decay products from two to ten, representative of the cases of two-body and multi-body decays of a $X$ particle with a uniformly distributed decay length $L_r$ in the range [0, 5]~m. An example of these simulated processes is depicted in Figure~\ref{fig:events}, where the case of two and ten tracks are reported. Each bin of the vertical axis corresponds to one of the 20 layers of the MDT chambers. The horizontal axis linearly maps the longitudinal coordinate of the MDT chambers. The number of bins on the horizontal axis is set to 333, a realistic average number of MDT tubes in the ATLAS detector. The images as in Figure~\ref{fig:events} constitute a convenient representation of the simulated and reconstructed physics process for training neural-network based algorithms. For each choice of charged particle multiplicity, 5k images are generated separately, with a  total of 45k available events. The sample is randomly split in two parts so that 80\% of the images are employed for the trainings and the remaining 20\% for the evaluations.

\section{Neural network models}\label{Sec:Models}
Algorithms for the trigger selection of the experiments at the LHC generically fall into two categories.
The first category is based on the ability to identify unique characteristics of the specific signature of interest, and defines a selection capable of preserving such a signature with high purity. This approach is effective for selected processes, but it lacks, by design, generalizability to other, possibly unknown, physics phenomena. The second category builds on the concept of anomaly detection~\cite{Collins:2018epr} to overcome the limitation of the first. In this scenario, the trigger selection is based on the likelihood that the event is not generated by known physics processes. This approach is particularly appropriate for searching for model-independent new physics signatures.

In this article two algorithms, representative of the two triggering philosophies just described, are developed and characterised.
A deep convolutional neural network (CNN) is trained for regressing the $L_r$ parameter of the neutral LLP while an autoencoder (AE) is trained exclusively on events where the decays of the LLP occurred near the interaction point for detecting anomalies. These two algorithms clearly follow the two distinct triggering criteria because the AE, contrarily to the CNN, is only exposed to events with short lifetimes in the training, hence remains agnostic on how a neutral LLP decay would look like in the detector. Once trained and deployed in the trigger and acquisition system, the CNN and the AE can be employed to define a selection criteria based, in the first case, on the inferred $L_r$ parameter and, in the second case, on the likelihood of the event to not only contain prompt decays.

\begin{figure}
\centering
\begin{subfigure}{1\textwidth}
\centering
\includegraphics[height=4.6cm]{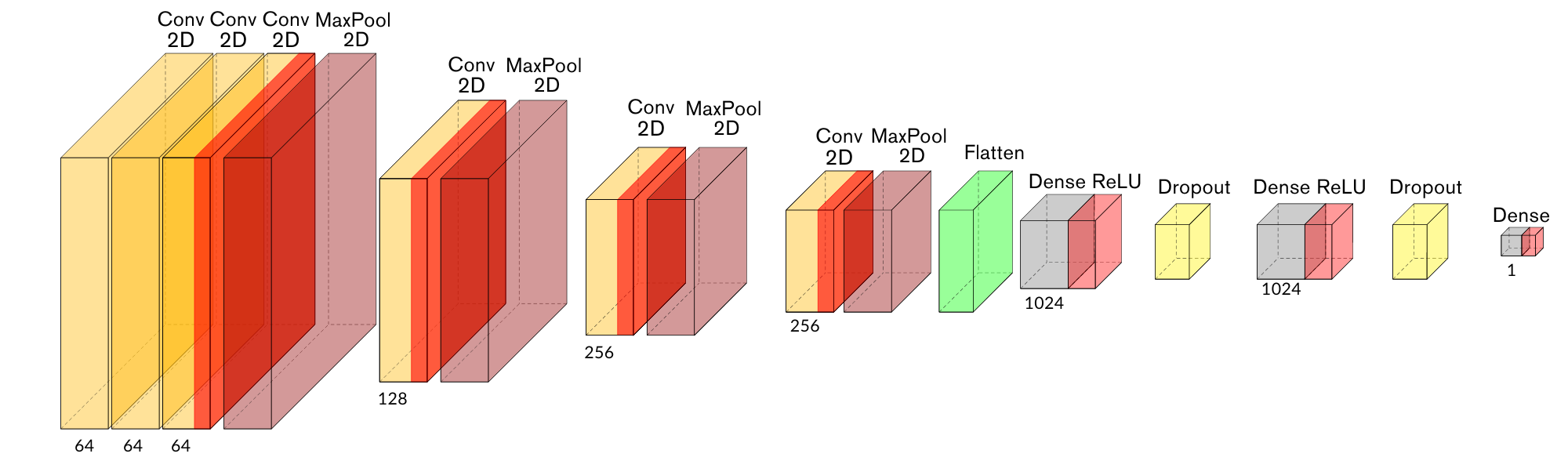}
\caption{CNN model}
\end{subfigure}
\hfill
\begin{subfigure}{1\textwidth}
\centering
\includegraphics[height=6cm]{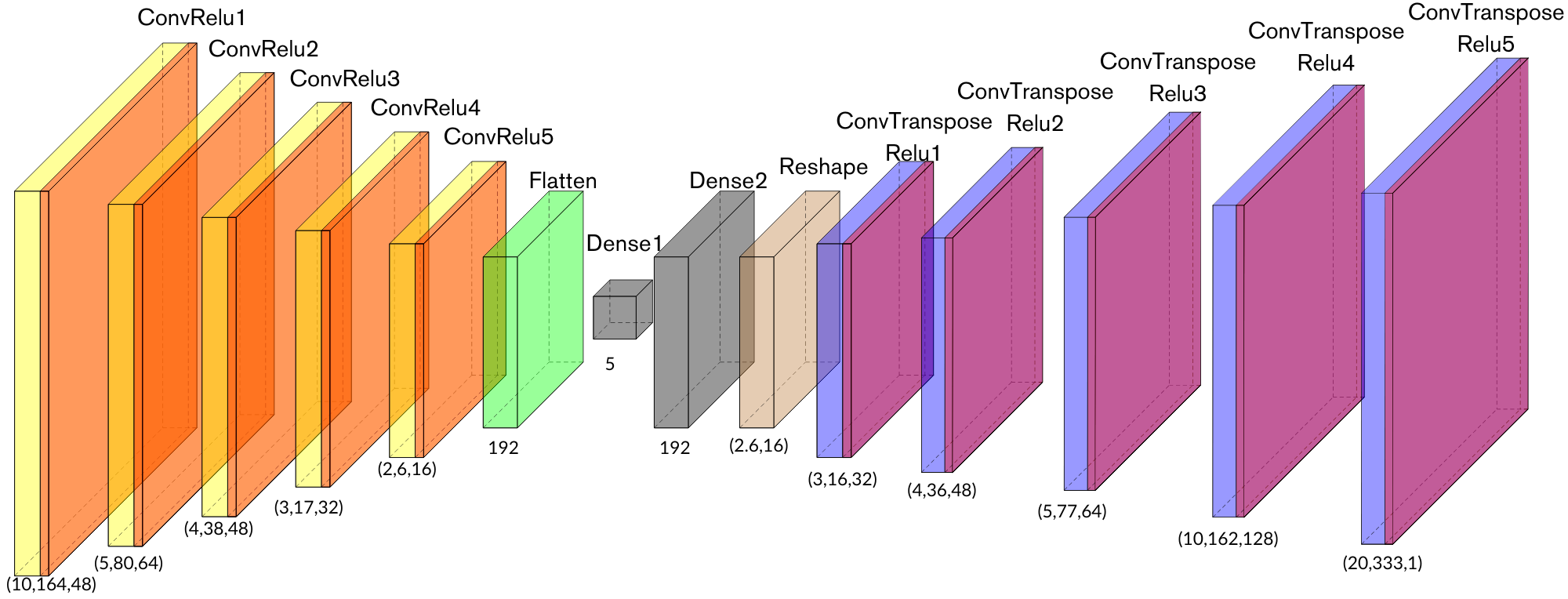} 
\caption{AE model}
\end{subfigure}
\caption{Diagrams with layer-by-layer details of the two architectures being considered in this work. The CNN model for regressing the $L_r$ parameter of the neutral LLP is in (a) while the AE model for detecting anomalies defined as decays not occurring near the interaction point is in (b). For both diagrams further details on the individual blocks of the diagram are given in the text.}
\label{fig:rete}
\end{figure}

The CNN model~\cite{Reference-CNN,deOliveira:2015xxd} is presented in Figure~\ref{fig:rete}. It comprises convolutional layers, ReLU activation functions, MaxPooling operators, and a final multi-layer perceptron with a single output node to regress the $L_r$ of the LLP. The implemented loss function is the mean squared error between the true and the predicted values, respectively $L_r$ and $\hat{L}_r$.

In a similar fashion, the AE is also based on a CNN and is also presented in Figure~\ref{fig:rete}. The encoder part is composed of convolutional layers, ReLU activations, and MaxPooling operators. The loss function of the AE is the binary cross-entropy and as such is responsible for the pixel-to-pixel comparison between the input and the reconstructed images. A second term to the loss was investigated but found to not provide substantial improvement in the discrimination performance. Such second term compared the high-level features in the latent space of another AE, constructed with the same architecture and trained on a different dataset with the same statistics, to compute a perceptual loss, inspired by the work in Ref.~\cite{perceptual-loss}. Once the AE is trained, only its encoder part is employed for constructing the discriminant for the trigger selection, as explained in Section~\ref{Sec:Results}, and consequently only the encoder is used when studying the performance and the inference time. For simplicity of convention, the encoder part of the AE is referred to the AE model throughout the text. The CNN model comes with~$\sim$2.8M trainable parameters, while the AE model with~$\sim$398k, and~$\sim$162k for the encoder part. The different number of parameters of the two models influences the studies on the inference time and throughput, as it will be shown in Section~4. We highlight how the chosen architectures are not ideal for the typical sparsity and cardinality of the data emerging from particle-physics collisions; they were chosen, instead, because they are fully supported by the adopted publicly available libraries. Support for other architectures, such as recurrent or graph neural networks, would definitely broaden the potential interest for the physics applications of fast inference on commercially available accelerator-based setups.

Data is labelled as background if the LLP decay is within $0$\,m~$<L_r<1$\,m and is labelled as signal if $3$\,m~$<L_r<5$\,m, and these definitions are consistently provided to both the CNN and AE models for the evaluation of the performances. Only background data is provided to the AE training, without any explicit label, while all the dataset, regardless of the truth $L_r$ value, is provided when training the CNN model. The CNN model includes data with decays to charged particles with multiplicity between two and ten, for both background and signal. Contrarily only decays with multiplicity between two and four for the background are considered when training the AE model. The reason is that the CNN model was found to provide excellent discrimination performance regardless of the composition of the background in terms of track multiplicity and opening angles of the decay products. Contrarily, the AE model was found to be more sensitive to the composition of the background and for this reason only the background with a low number of tracks was kept in the training. Further optimisation of the AE architecture and training procedure, and also proper simulations of background events originating from SM particles decaying promptly in the detector, are possible and are left for future studies.

For studying the inference time two different FPGA boards are considered, the Xilinx Alveo U50 and U250. The AE model was not run on the U250 board due to missing support of the Xilinx Vitis-AI tool to the Reshape layer, which is included in the architecture of the AE. It is also important to note that the servers hosting the two accelerator cards are equipped differently, hence a direct comparison between the performance of the U50 and U250 cards for the CNN model is not straightforward given the different CPU load on each. 

The actual acceleration in the FPGA cards requires several preliminary operations and the Xilinx distributed Vitis-AI tool provide a complete workflow for this purpose. The Post Training Quantization of Vitis-AI is chosen for the model quantization, converting the original trained parameters to 8-bit integer precision. With the quantized model the Vitis-AI compiler allows to have the so-called Xilinx Intermediate Representation (XIR) of the model, a representation of the model in a workable format for the specific chosen FPGA board. Vitis-AI provides a Python-based script which manages the model inference using XIR and Vitis AI Run Time libraries operations. The first part of the script consists in the compiled model deserialization, where, starting from the model XIR representation, a \textit{xmodel} file is returned. Then a final utility checks the correctness of the XIR model and performs the actual deployment to the accelerator card.

The two Alveo cards come with different architecture configurations and also contain different Deep-Learning Processing Units (DPUs). In addition, slightly different quantization bit-width methods are employed, together with two different versions of the Vitis-AI software, v1.4.1 and v2.5, respectively for the U50 and U250 boards.

\section{Results}\label{Sec:Results}
In this section, we present the performance evaluation of both the CNN and AE models in terms of performance, inference time and throughput.

\begin{figure}[t!]
\centering
\begin{subfigure}{0.49\textwidth}
\centering
\includegraphics[width=8cm]{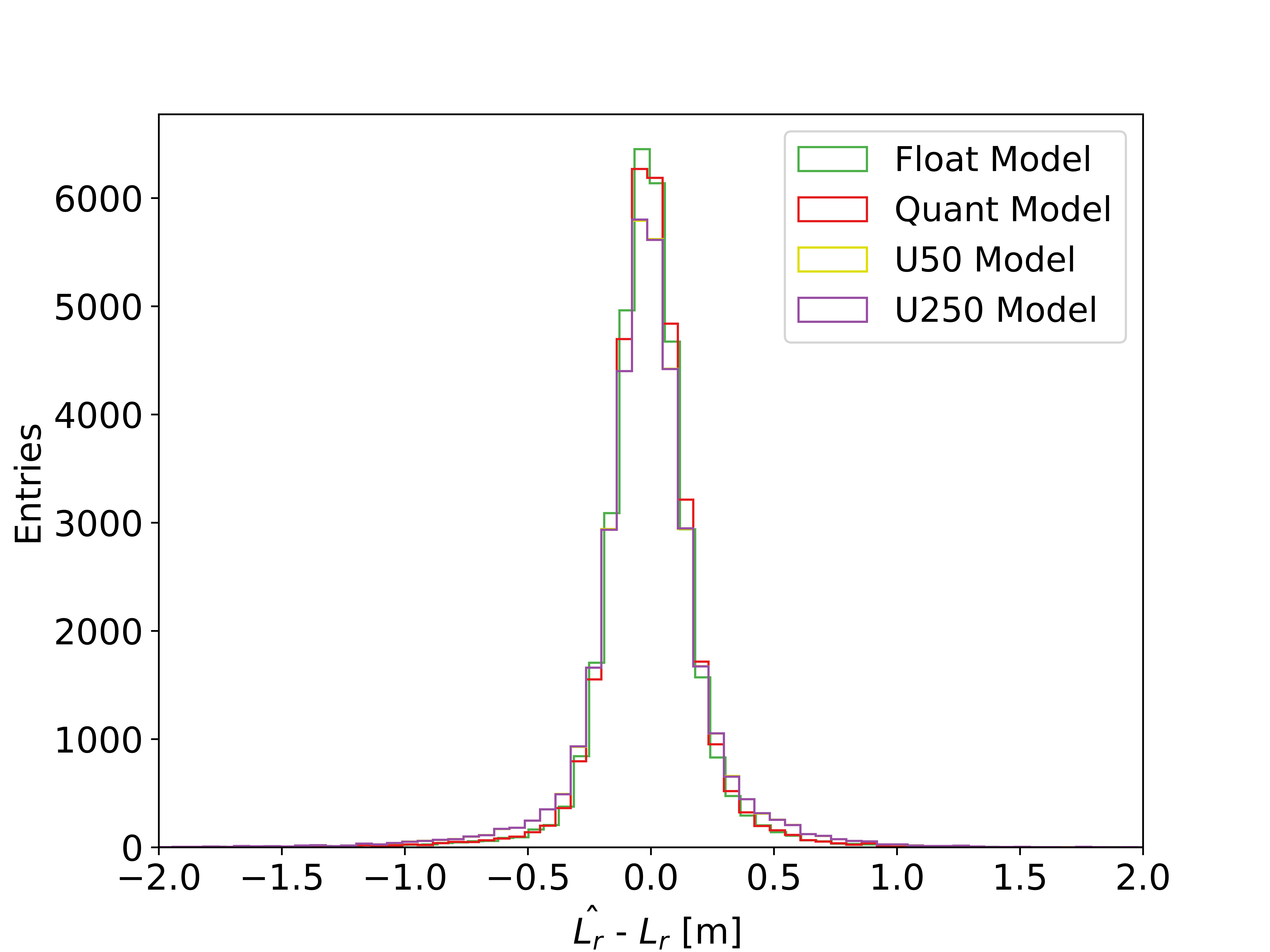}
\caption{}
\end{subfigure}
\hfill
\begin{subfigure}{0.49\textwidth}
\centering
\includegraphics[width=8cm]{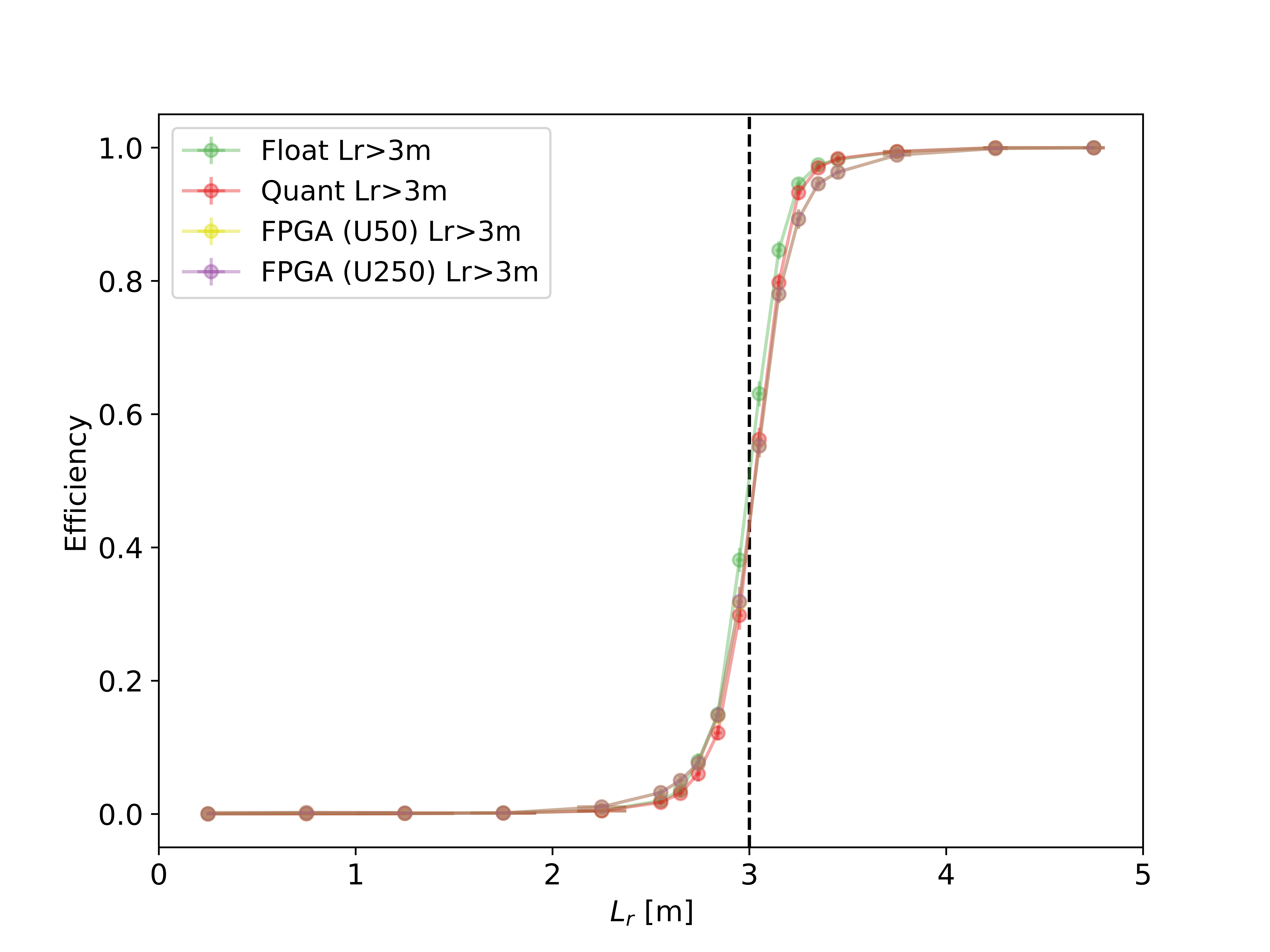}
\caption{}
\end{subfigure}
\caption{(a) Residual plot between the true decay length of the neutral LLP and the one predicted by the CNN model. (b) Efficiency plot of the CNN model, where the efficiency is defined as the fraction of neutral LLP decays with the predicted decay length $\hat{L}_r>3$\,m as a function of the true decay length $L_r$.}
\label{fig:regressor}
\end{figure}

\begin{figure}[h!]
\centering
\includegraphics[width=9cm]{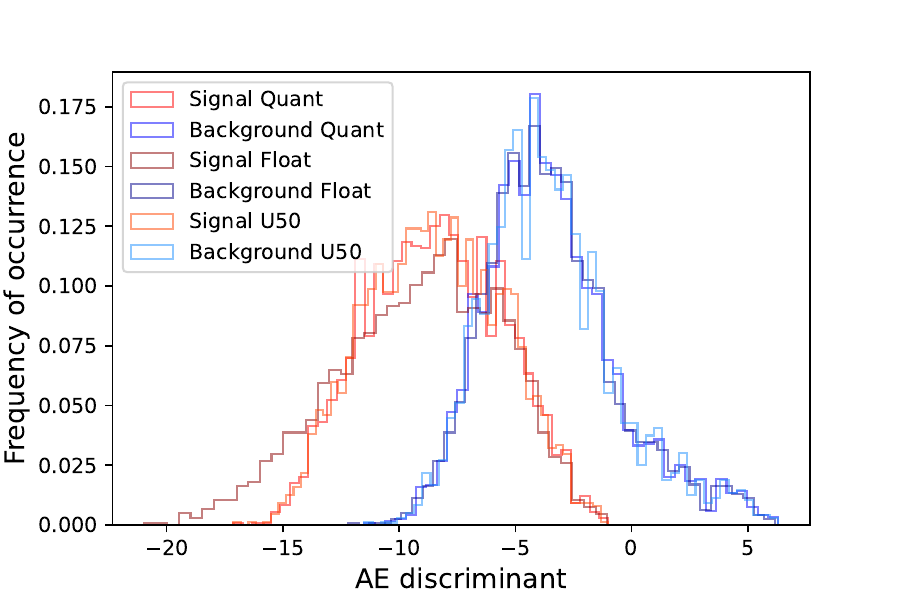}
\caption{Discriminant of the AE model, defined as the sum of the hidden features in the latent space, for signal and background images evaluated for the float and quantized models and the one deployed on the Xilinx Alveo U50 accelerator card. Identically to the CNN model, LLP decays are labelled as signal if $3$\,m~$<L_r<5$\,m and as background if $0$\,m~$<L_r<1$\,m.}
\label{fig:binary}
\end{figure}

\begin{figure}[t!]
\centering
\begin{subfigure}{0.49\textwidth}
\centering
\includegraphics[width=8cm]{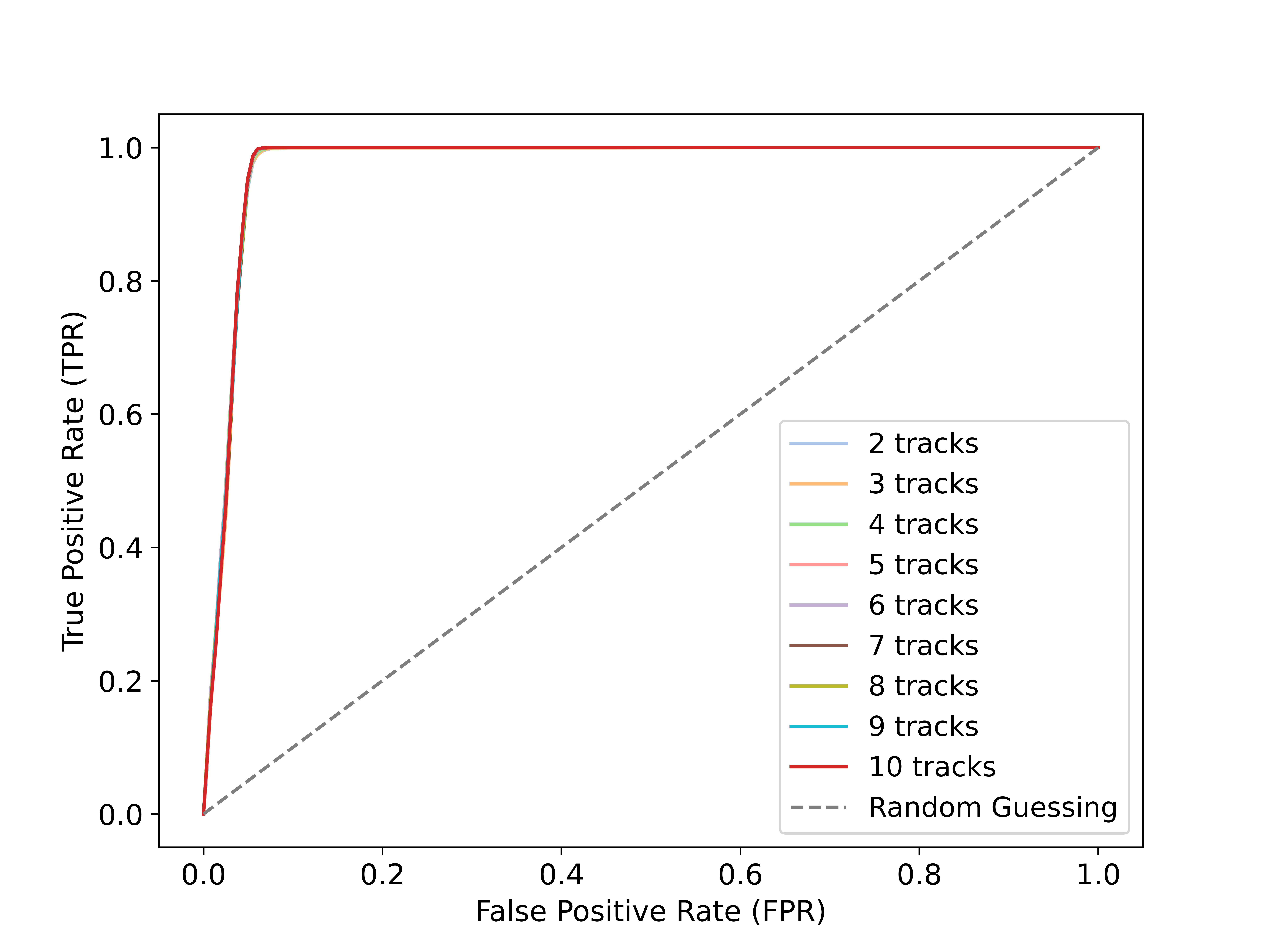}
\caption{CNN model}
\end{subfigure}
\hfill
\begin{subfigure}{0.49\textwidth}
\centering
\includegraphics[width=8cm]{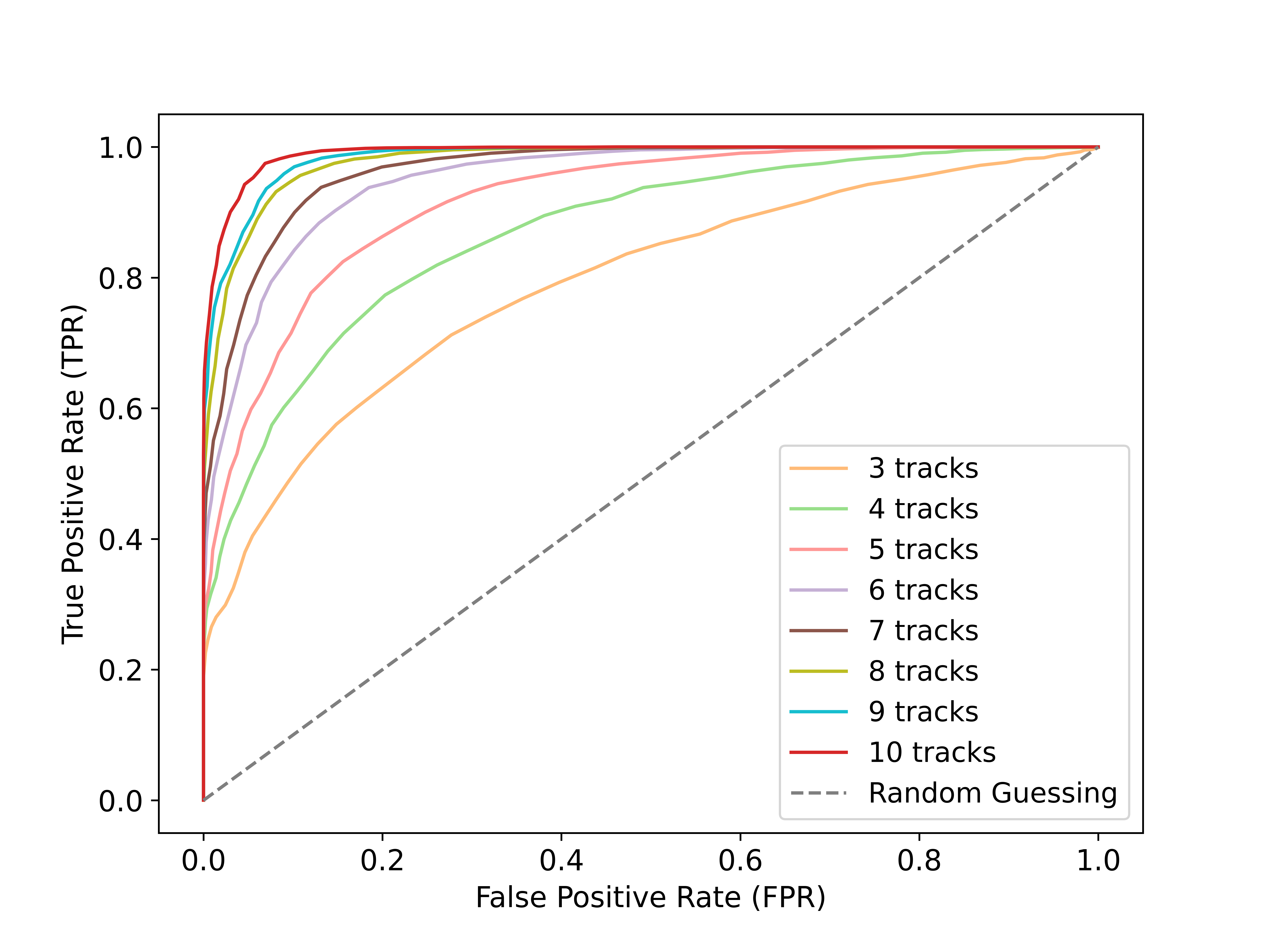}
\caption{AE model}
\end{subfigure}
\caption{ROC curves for the (a) CNN and the (b) AE models. In both cases, LLP decays are labelled as signal if $3$\,m~$<L_r<5$\,m and as background if $0$\,m~$<L_r<1$\,m. Track multiplicity between two and ten, and between two and four, is used for creating the background dataset, respectively for the CNN and the AE models. In contrast, the track multiplicity is considered separately for signal, as a way to estimate the discrimination performance for different hypotheses of new physics signatures. The CNN model is found to provide better discrimination performance than the AE model, and to not be dependent on the track multiplicity of the neutral LLP decay.}
\label{fig:roc}
\end{figure}

Figure~\ref{fig:regressor} presents the distribution of the residuals between the predicted and the true decay length of the neutral LLP and the trigger efficiency of the CNN model, considering the float, the quantized, as well as the models actually deployed on the U50 and U250 accelerator cards. The quantization of the models resulted in a small degradation in accuracy but did not significantly impact the efficiency curve, indicating an acceptable  performance degradation. Similarly, a slight degradation was observed between the quantized model and the one deployed on the FPGA. The trigger efficiency is defined as the fraction of neutral LLP decays with the predicted decay length $\hat{L}_r>3$~m as a function of the true decay length $L_r$. The steep turn-on curve of the efficiency confirms the residual distributions being under control and indicates the feasibility of constructing a trigger-like selection able to provide a sample enriched with signals and depleted in background events.

The evaluation results of the AE are shown in Figure~\ref{fig:binary} with a discriminant defined as the sum of the hidden features of the AE latent space. Other variations of the discriminant were investigated but due to the sparsity and the relative contribution of noise in the original and reconstructed images, a discriminant based on the latent space features was found to be more adequate for the purpose of anomaly detection. Among the possible choices for combining the latent space features the sum was considered for its simplicity and compatibility with the Vitis-AI software. As with the CNN model, a small degradation is observed when quantizing the model while the quantized model and the one deployed on the FPGA were found in substantial agreement.

\begin{table}[t!]
\centering
\begin{tabular}{lllll}
& \cellcolor[HTML]{C0C0C0}CPU & \cellcolor[HTML]{C0C0C0}GPU & \cellcolor[HTML]{C0C0C0}U50 & \cellcolor[HTML]{C0C0C0}U250 \\
\cellcolor[HTML]{EFEFEF}Inference time {[}ms{]} & 5.1 $\pm$ 1.1 & 1.0 $\pm$ 0.1 & 3.7 $\pm$ 0.1 & 3.1 $\pm$ 0.4 \\
\cellcolor[HTML]{EFEFEF}Throughput {[}fps{]} & 302 $\pm$ 4 & 9930 $\pm$ 187 & 950 $\pm$ 5 & 553 $\pm$ 4 \\
\end{tabular}
\caption{Inference time in ms and throughput in frames per second for the CNN model on different target architectures. The results include the actual deployment of the model on the FPGA U50 and U250 accelerator cards.}
\label{tab:time-CNN}
\vspace{1cm}
\begin{tabular}{lllll}
& \cellcolor[HTML]{C0C0C0}CPU & \cellcolor[HTML]{C0C0C0}GPU & \cellcolor[HTML]{C0C0C0}U50 \\
\cellcolor[HTML]{EFEFEF}Inference time {[}ms{]} & 0.7 $\pm$ 0.1 & 0.41 $\pm$ 0.01 & 2.6 $\pm$ 0.3 \\
\cellcolor[HTML]{EFEFEF}Throughput {[}fps{]} & 3477 $\pm$ 210 & 79238 $\pm$ 2358 & 1497 $\pm$ 3 \\
\end{tabular}
\caption{Inference time in ms and throughput in frames per second for the AE model on the different target architectures. The results include the actual deployment of the model on the FPGA U50 accelerator card.}
\label{tab:time-AE}
\end{table}

The performances of the two models are also studied with the ROC curves presented in Figure~\ref{fig:roc}. These curves are computed by labelling consistently as signal the LLP decays with $3$\,m~$<L_r<5$\,m, and as background those decays with $0$\,m~$<L_r<1$\,m. The background with the different charged particle multiplicities is summed up, and ROC curves are constructed with respect to a signal with a particular number of charged particle multiplicity. To be consistent with the training procedure outlined in Section~3, track multiplicity of the neutral LLP decay between two and ten, and between two and four, is used for creating the background sample, respectively for the CNN and the AE models.
The ROC curves demonstrate the capability of the CNN model to effectively learn the decay position independently of the multiplicity of the charged decay products, while a dependence on the multiplicity is clearly evident for the AE model. The performance of the AE model in case of two-track signals is not displayed because no discrimination is achieved in this case. The performance of the AE model was also found to be dependent on other aspects of the generation, for example on the opening angle of the decay productions of the neutral LLP. Additional studies in this direction are considered out of the scope of this article, as these results discussed so far already demonstrate the capability of the chosen network architectures to effectively select the neutral LLP decays of interest.

\begin{figure}[t!]
\centering
\includegraphics[width=9cm]{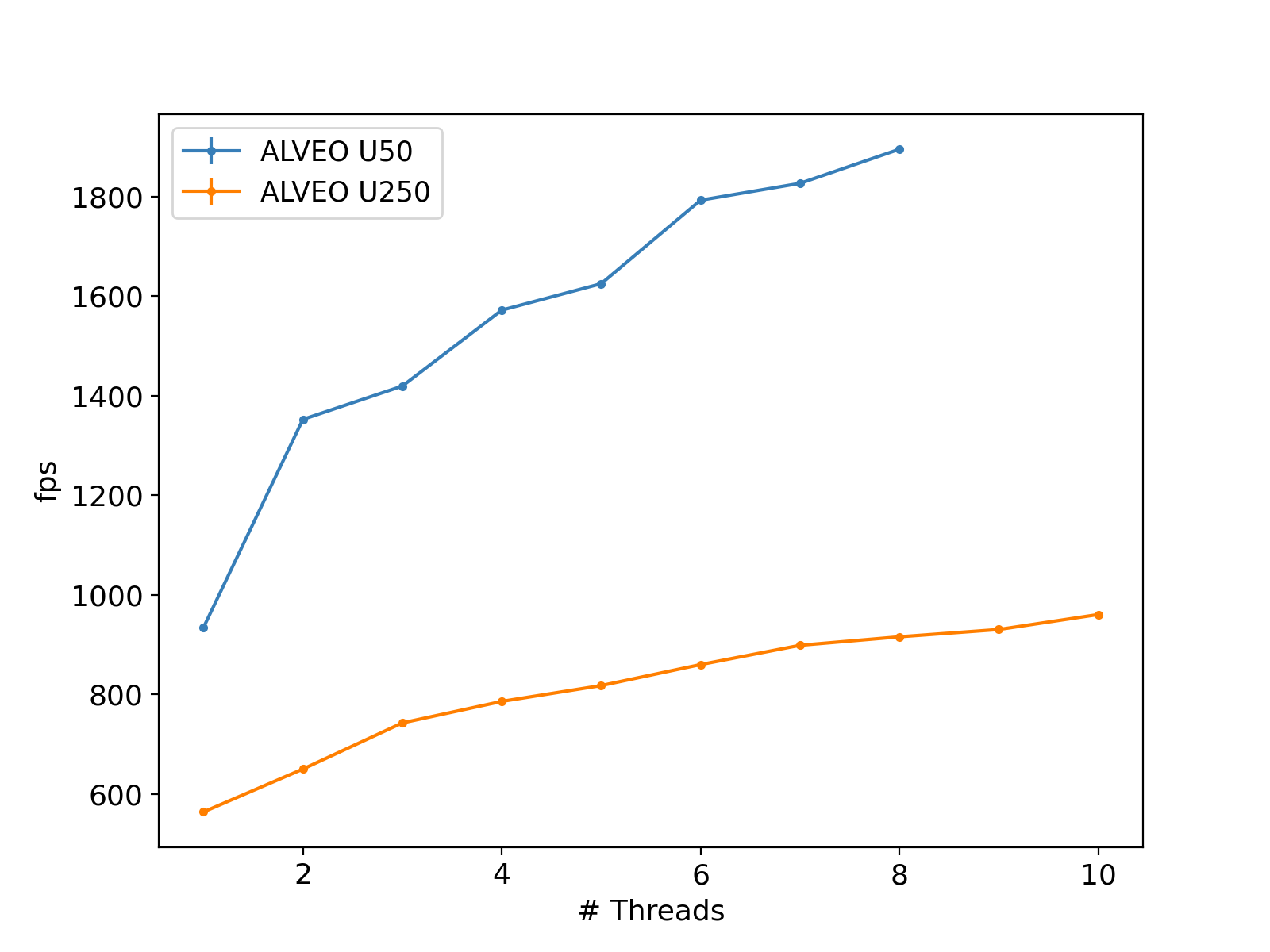}
\caption{Throughput in frames per second as a function of the number of concurrent threads as measured with the CNN model deployed on the FPGA U50 and U250 accelerator cards.}
\label{fig:FPGA-threading}
\end{figure}

The inference time and the throughput of the CNN and AE models on different architectures are also studied and results are presented in Tables~\ref{tab:time-CNN} and \ref{tab:time-AE}. The inference on the FPGA accelerator cards require to batch images with a fixed size, which depends on the DPU of the accelerator card and is declared by the manufacturer. The U50 and U250 cards require a batch size of three and four, respectively. For consistency in the reported results, a batch size of four is also implemented for studying the inference time and the throughput on CPU and GPU architectures. The measurements on the CPU and GPU architectures have been performed by converting the model into the Open Neural Network Exchange (ONNX) format with the runtime engines corresponding to these two architectures~\cite{ONNX}. The ONNX format was found to substantially improve the results, even more than one order of magnitude on both architectures. The measurements on the CPU were performed using all the cores and on a machine equipped with AMD EPYC 7302 16-Core processors. The measurements on the GPU were performed on a GPU NVIDIA Tesla V100, and using the float models before quantization. The inference time results are obtained by averaging on few tens of measurements. The throughput is estimated by inferring the models with 10k images. The first measurements of both inference time and throughput on accelerators are discarded since they were observed to be systematically higher.

Overall the study indicates that all architecture technologies offer inference time and throughput adequate for the typical latency requirements of a high-level trigger selection in a general-purpose experiment at LHC or HL-LHC. The inference time for the CNN model suggests that the acceleration on FPGA gives an advantage compared to the CPU-based approach. A similar advantage is not evident for the AE model. This can be attributed to the lightness of the model in terms of number of parameters, which results in the actual inference time being negligible compared to the time needed for loading the data onto the FPGA itself. A proper study of the time needed for performing the various sub-tasks for enabling the inference on the FPGA is considered of great interest, but was not possible given the provided tools at hand. The throughput measurements also indicate the superiority of the FPGA-acceleration approach compared to the CPU-based one for the CNN model, and not for the AE model for the same considerations just expressed. In addition the throughput on the GPU architecture seems to suggest the superiority of this approach but this is achieved, as the corresponding measurements on the inference time confirm, only thanks to the capability of GPUs to process inference concurrently, and such high degree of concurrent computing can't be directly injected within a multi-node high-level trigger farm at colliders.

In summary, considering the necessity of deploying larger models when dealing with real experiments, the results presented in this study suggest that a heterogeneous computing model with FPGA-based acceleration has the potential to improve the real-time processing and the responsiveness of a trigger system. It is also worth noting that the inference time and throughput for the Xilinix Alveo U50 and U250 FPGAs, as presented in Tables~\ref{tab:time-CNN} and~\ref{tab:time-AE}, were obtained without utilizing the device multi-threading capability. We tested the performance in terms of throughput with varying numbers of parallel threads, and observed an almost linear improvement in the throughput performance versus the number of concurrent threads, as show in Figure~\ref{fig:FPGA-threading}. Hence, by leveraging the multi-threading capabilities, it becomes possible to achieve superior performances compared to what shown in Tables~\ref{tab:time-CNN} and~\ref{tab:time-AE}. One final consideration is the comparison of the power dissipation declared by the manufacturers for the considered architectures. The NVIDIA Tesla V100 GPU has a power consumption of 300~W, to be compared with the Xilinx Alveo U50 of 75~W. The actual power dissipation will obviously depend on the amount of resources being used in reality when performing the inference. This has not been studied because it will depend on the multi-node architecture of the farm equipped or not equipped with accelerators, hence what is declared by the manufacturers is considered of enough interest for corroborating these remarks. 
More studies in this direction will be necessary and are considered an important step for the definition of the trigger and data acquisition system of the future experiments operating at the HL-LHC.

\section{Conclusions}\label{Sec:Conclusions}
This article discusses the performance evaluation of machine learning algorithms on commercially available Xilinx FPGA accelerator cards. The necessary steps including model training, quantization, compilation, and actual deployment on the FPGA board were all performed. The post-training quantization technique provided by Vitis-AI was used for model quantization, which resulted in an acceptable level of model accuracy degradation and reduced model size. Two neural-network models based on different architectures were trained and characterised for selecting events with neutral long-lived particle decays within the geometrical acceptance of a muon spectrometer of a general-purpose experiment at the LHC. A model based on convolutional neural networks and trained to regress the decay length of the neutral long-lived particle and a model based on an auto-encoder architecture and trained to detect as anomalies such decays are presented. The first model was deployed on Xilinx Alveo U50 and U250 accelerator cards using the Vitis-AI compiler, while the second model was only deployed on the U50 card. The models were demonstrated to efficiently retain events of physics interest while rejecting background collisions. The inference time and throughput of the models were also confronted on a CPU and on different architectures with GPU-based or FPGA-based acceleration. The results indicate that the measured inference times on all tested architectures fit within the typical latency requirements of a high-level trigger selection in a general-purpose experiment at LHC or HL-LHC. 

\section{Acknowledgments}
We thank the INFN IT teams in Genoa and Rome, and in particular Mirko Corosu and Luca Rei, for useful support in instrumenting the local computing resources to accommodate the FPGA accelerator cards. This work is partially supported by ICSC -- Centro Nazionale di Ricerca in High Performance Computing, Big Data and Quantum Computing, funded by European Union -- NextGenerationEU.

\end{document}